\documentclass[12pt]{iopart}

\usepackage{graphicx}
\usepackage{subfigure}

\newcommand{\gen}{\textrm{GeV}}
\newcommand{\ten}{\textrm{TeV}}

\newcommand{\gmom}{\textrm{GeV}/c}
\newcommand{\mmass}{\textrm{MeV}/c^2}

\newcommand{\phir}{\varphi(1020)}
\newcommand{\kstar}{\textrm{K}^*(892)^0}
\newcommand{\sstar}{\Sigma(1385)^{\pm}}

\newcommand{\xstar}{\Xi(1530)^0}
\newcommand{\kpi}{\textrm{K}\pi}
\newcommand{\pT}{p_{\textrm{t}}}

\begin{document}

\title[Resonances in ALICE]{Resonance measurements in pp and Pb--Pb collisions with the ALICE detector}

\author{A. Pulvirenti, for the ALICE Collaboration}
\address{INFN Sezione di Catania, Via S. Sofia 64, Catania (Italy)}
\ead{alberto.pulvirenti@ct.infn.it}

\begin{abstract}
The study of resonance production in pp collisions helps understanding hadronization mechanisms and
tuning the QCD-inspired particle production models.
In Pb--Pb collisions, resonances allow one to probe the temperature and time evolution of the fireball.

Transverse momentum spectra have been analyzed for $\kstar$, $\phir$ and $\xstar$ resonances using data from 
pp collisions at 7 TeV collected by the ALICE detector. 
A comparison with Monte Carlo event generators shows different levels of agreement for meson spectra,
while $\xstar$ is always underestimated.
\end{abstract}



\section{Introduction}

The study of resonance production plays an important role both in pp and heavy-ion high energy collisions.
In pp collisions, it contributes to the general understanding of hadron production mechanisms and 
provides a reference for tuning QCD-inspired models.

In heavy-ion collisions, due to their short lifetime (few fm/$c$), resonances can decay inside the fireball,
and studying the interactions of their decay products with the surrounding medium allows one to estimate its
temperature and investigate its time evolution~\cite{torrieri_rafelski}.
Moreover, resonances produced in the early stages of the collision can show signals of chiral symmetry restoration 
in terms of modifications of their mass and/or width~\cite{rapp_wambach}.

ALICE~\cite{ALICE_JINST} is the LHC experiment mainly devoted to the study
of ultra-relativistic heavy-ion collisions, in order to investigate the properties of the Quark-Gluon Plasma (QGP).
Its excellent capabilities for particle identification~\cite{kalweit} over a wide $\pT$ range 
(from $\sim$0.1 to $\sim$10~$\gmom$), make it a very well suited device for the study of resonances.

Results are presented for $\kstar$ (identified by charged $\kpi$ decay channel), $\phir$ (K$^+$K$^-$ decay channel)
and $\xstar$ ($\Xi\pi$ decay channel) measurements done with a sample of data collected by ALICE in 2010.
Section~\ref{sec:analysis} briefly describes the analysis procedure and Section~\ref{sec:results}
shows some results. Section~\ref{sec:conclusions} presents the conclusions and an outlook.


\begin{figure}
   \centering
   \subfigure{\includegraphics[height=40mm]{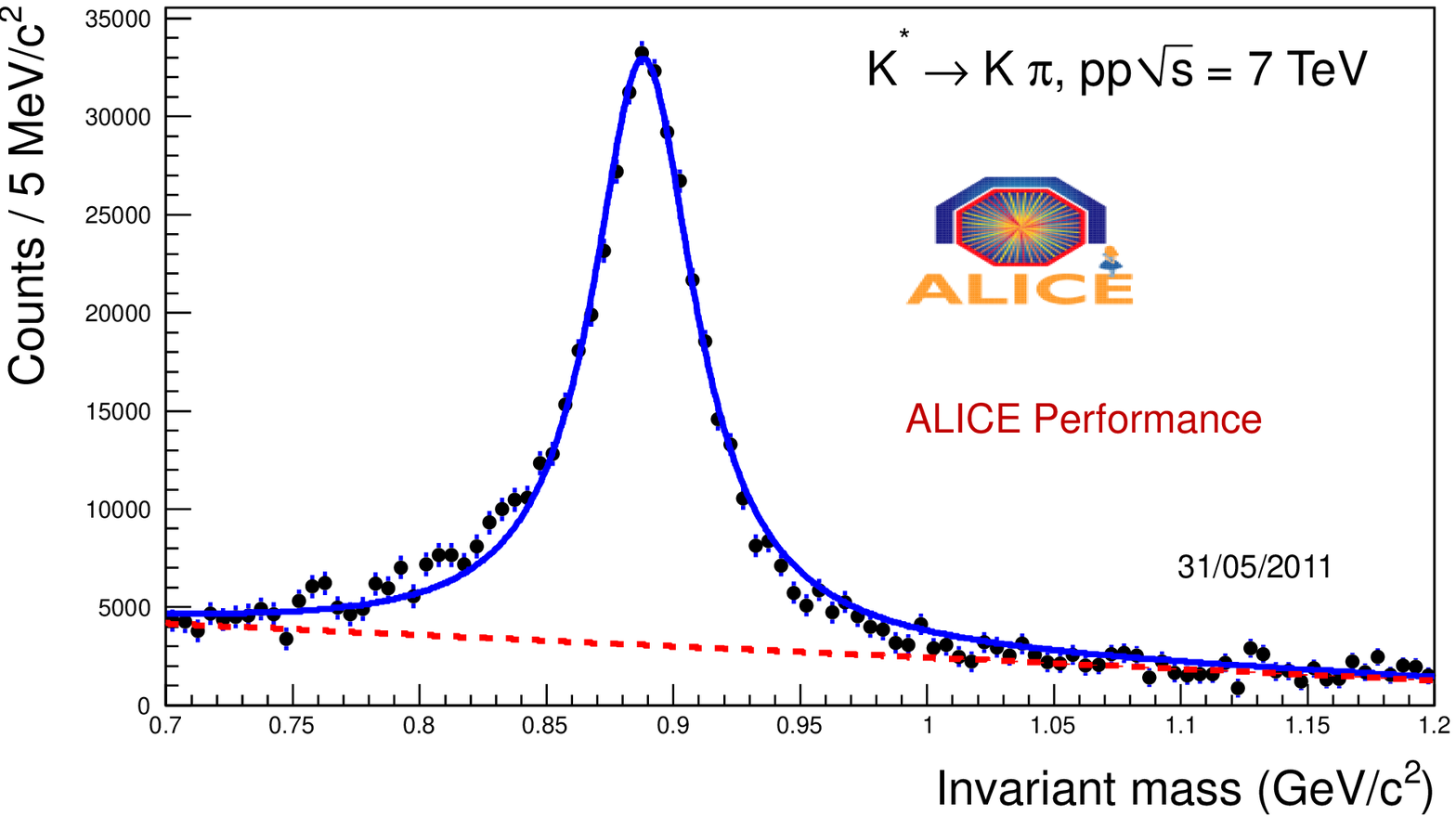}}
   \subfigure{\includegraphics[height=40mm]{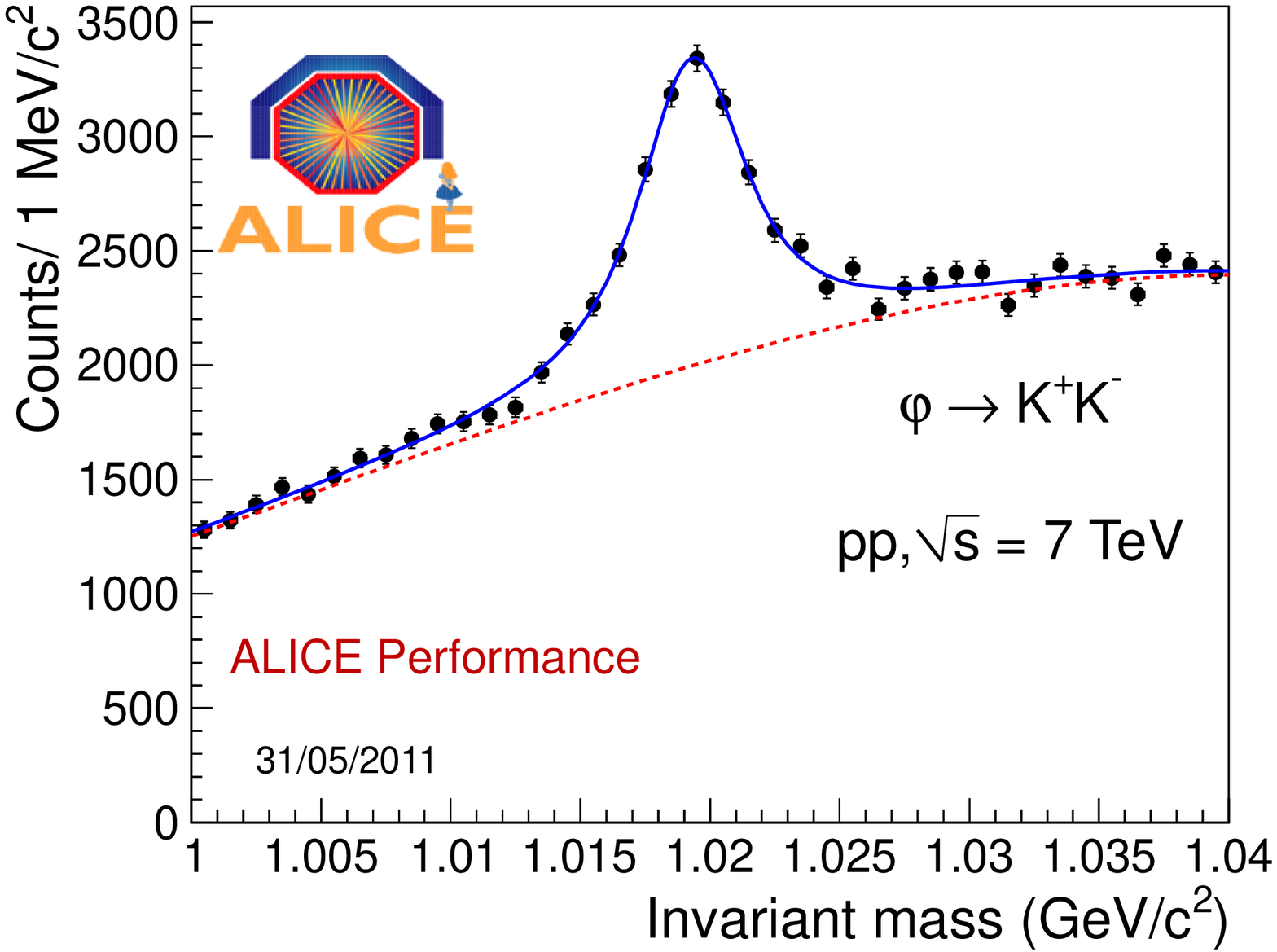}}
   \caption{Left: invariant mass distribution of charged $\kpi$ pairs in the $\pT$ bin
            $2.0\leq\pT\leq 2.5$~$\gmom$.
            Solid line is the fit function, consisting in a Breit-Wigner on a linear background.
            Dashed line represents the background.\\
            Right: invariant mass distribution of K$^+$K$^-$ pairs in the $\pT$ bin $0.7\leq\pT\leq0.8$~$\gmom$.
            Solid line is the fit function, consisting in a Voigtian plus a 3$^{\textrm{rd}}$ order polynomial.
            Dashed line represents the background.}
   \label{fig:peaks}
\end{figure}
\begin{figure}
   \centering
   \subfigure{\includegraphics[width=75mm]{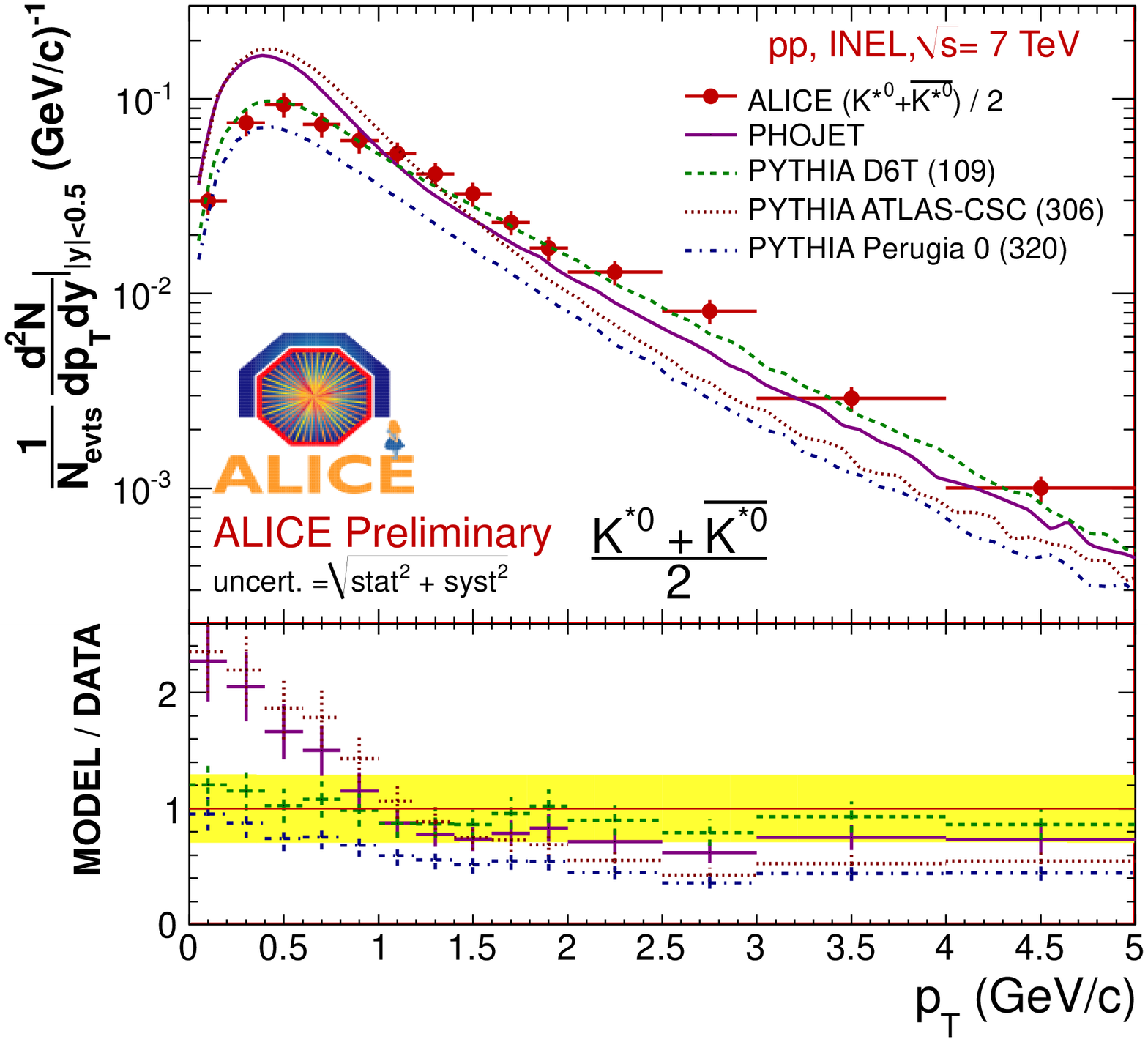}}
   \vspace{1mm}\subfigure{\includegraphics[width=75mm]{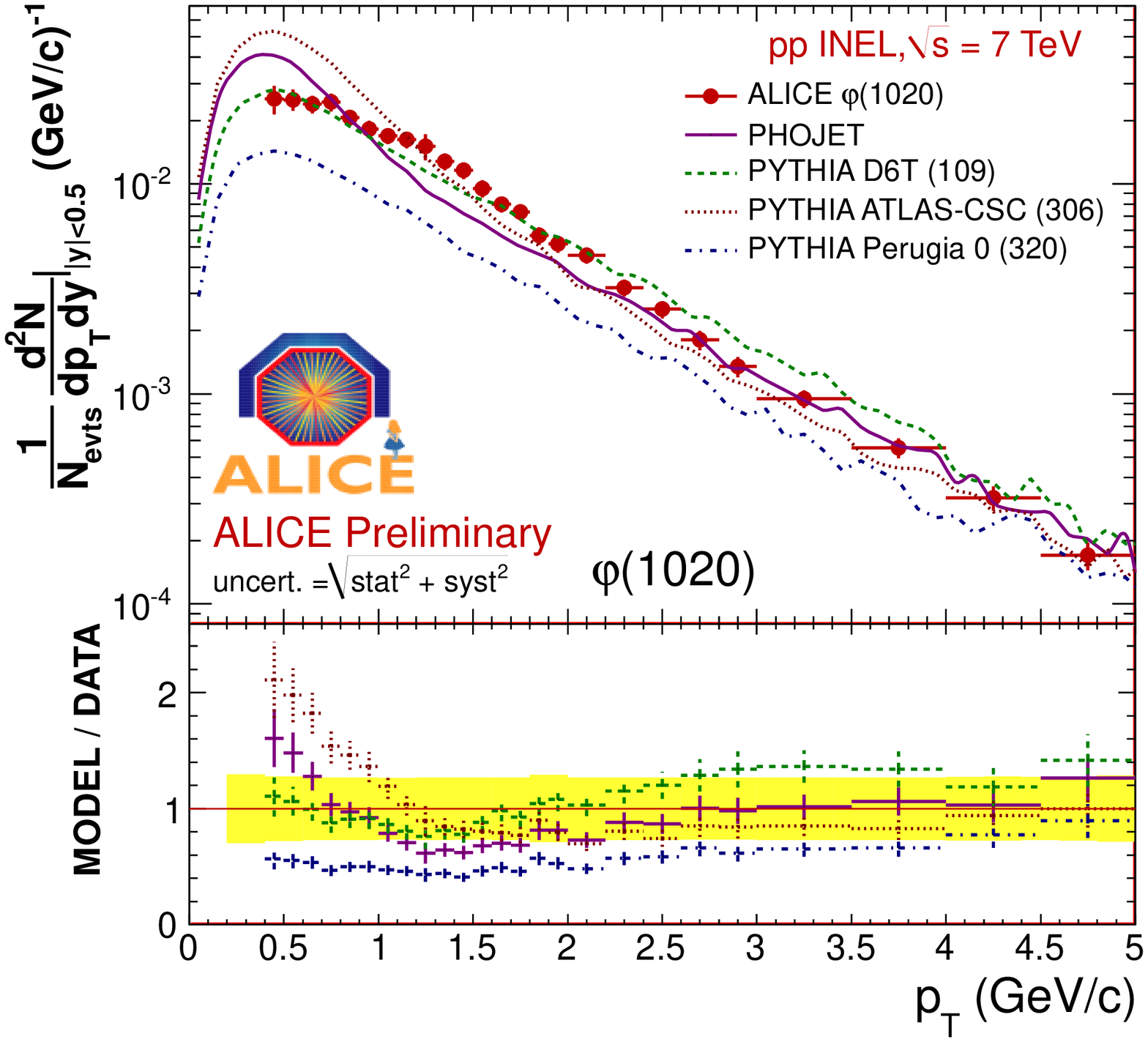}}
   \vspace{1mm}\subfigure{\includegraphics[width=75mm]{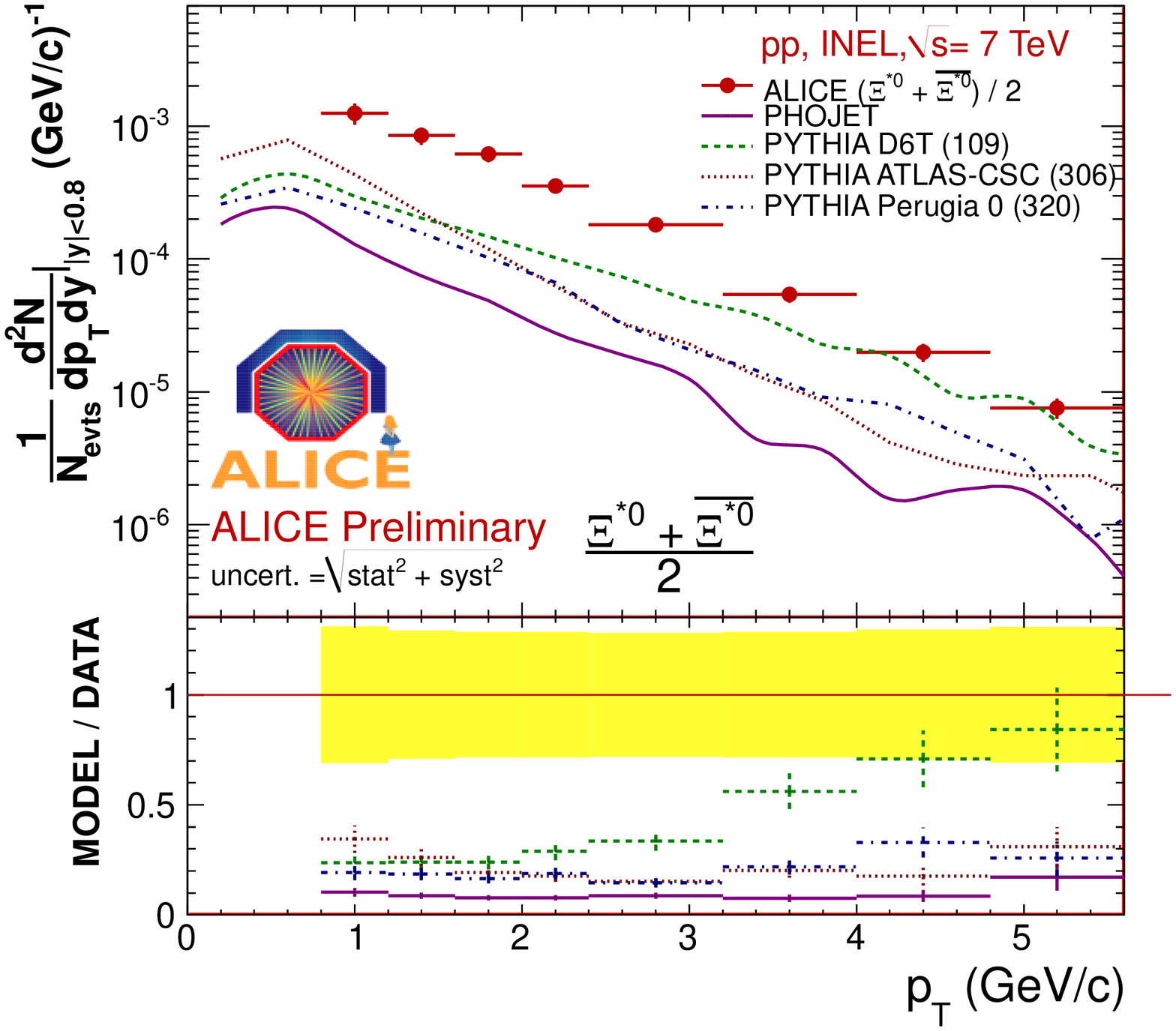}}
   \caption{(Color online) Spectra of $\kstar$, $\phir$ and $\xstar$ compared with Monte Carlo.
            Filled circles are data, curves describe the used generators (see text). 
            Lower part of each plot shows the ratio between Monte Carlo simulation and data.
            The shadow band corresponds to the systematic errors on the data.}
   \label{fig:spectra}
\end{figure}
\begin{figure}
   \centering
   \subfigure{\includegraphics[width=70mm]{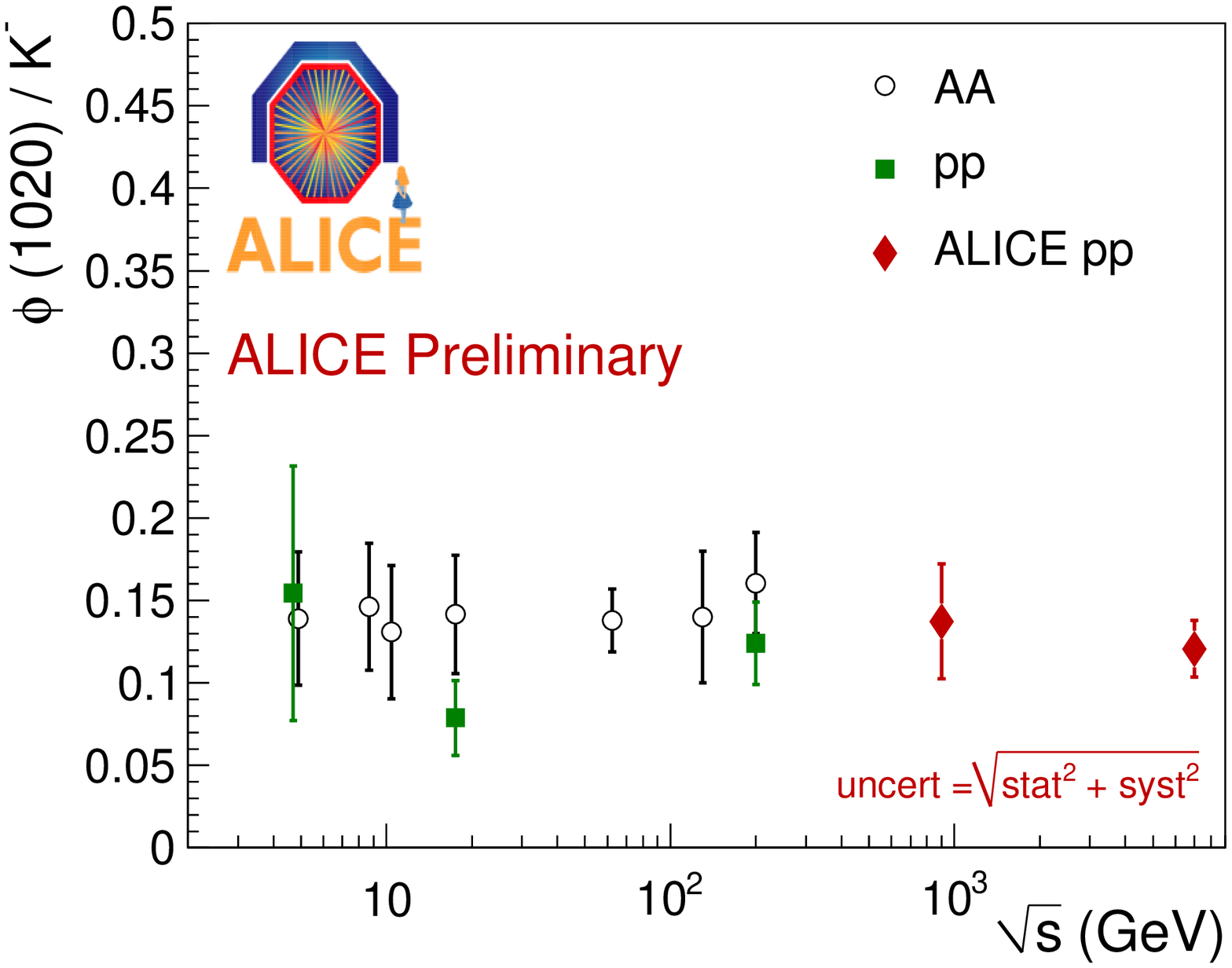}}
   \subfigure{\includegraphics[width=70mm]{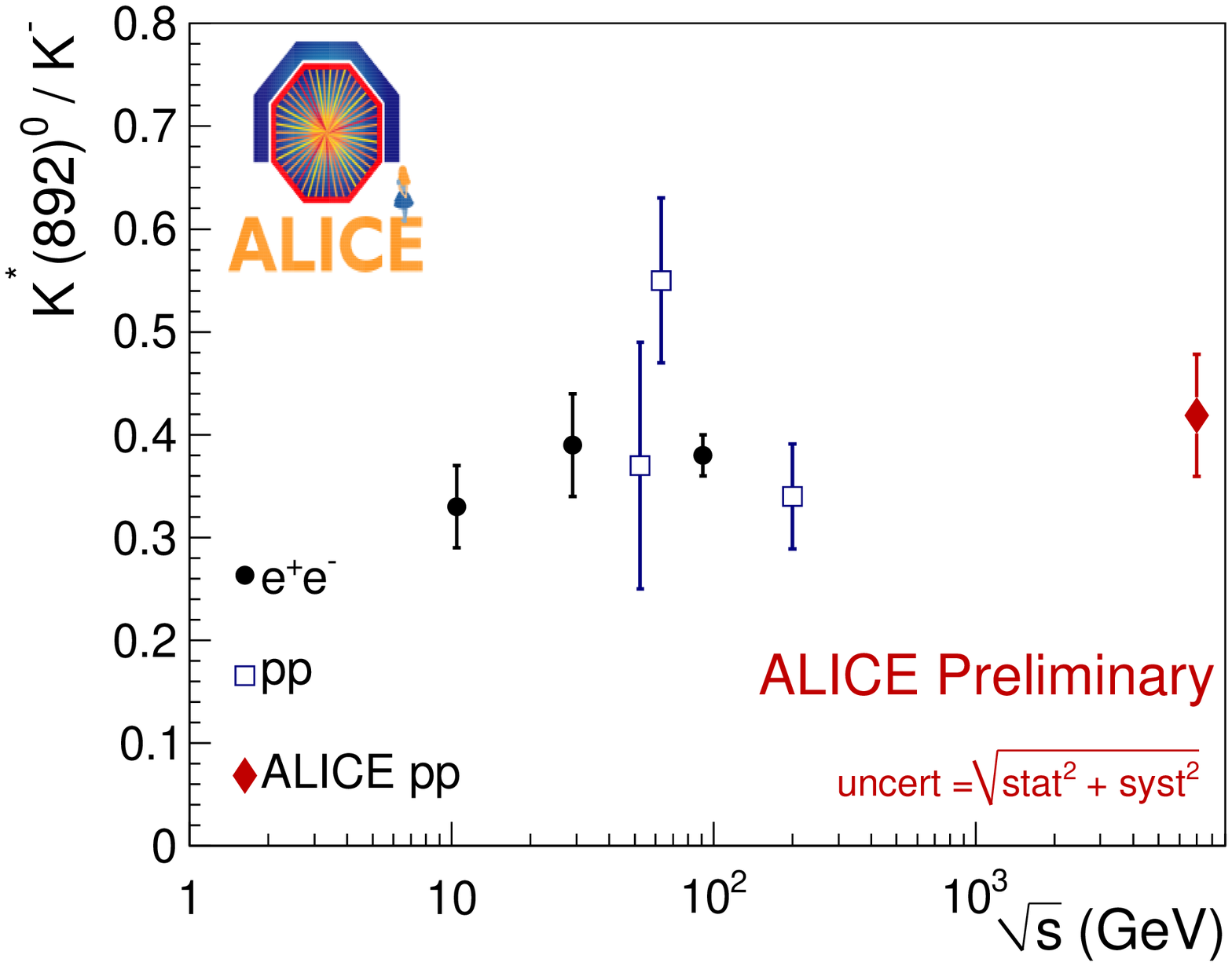}}
   \caption{(Color online) Resonance to stable particle ratios as a function of collision energy~\cite{ratios1,ratios2}.
            Diamonds represent ALICE measurements at 7 $\ten$ and, when available, at 900 $\gen$~\cite{paper900gev1,paper900gev2}.}
   \label{fig:ratios}
\end{figure}
%

\section{Analysis procedure}\label{sec:analysis}

The data analysis is carried out using a sample of minimum-bias pp data collected by ALICE during 2010, with 
a size ranging from 25 to 155 million events, for the different resonances analyzed.
In order to minimize the background, particle identification selection criteria for the decay products are used, which are based on their d$E/$d$x$ measured
by the Time Projection Chamber (TPC) and their $\beta$ measured by the Time of Flight (TOF) detector.

Due to their very short lifetime, resonance decay products cannot be distinguished from the particles coming from the primary vertex,
and their yield can only be measured by computing the invariant
mass spectrum of all primary tracks and then subtracting the combinatorial background.
For the $\kstar$, the combinatorial background is estimated from like-sign $\kpi$ pairs and then subtracted from
the unlike-sign $\kpi$. 
The resulting distribution is fitted with a Breit-Wigner function plus a straight line 
to account for a remaining residual background (figure~\ref{fig:peaks} left).
The raw yields of $\phir$ and $\xstar$ are extracted by fitting the invariant mass distributions with a function composed by a polynomial 
to describe the background, plus a Voigtian, i.e. the convolution of a Gaussian and a Breit-Wigner (figure~\ref{fig:peaks} right).
This choice is necessary to take into account the resolution in invariant mass, 
since it ranges between 1 and 2~$\mmass$ and is therefore comparable with the nominal widths of 
these resonances (4.25 and 9.1~$\mmass$, respectively).


\section{Results}\label{sec:results}

The raw yields of $\kstar$, $\phir$ and $\xstar$ in pp collisions are corrected for reconstruction efficiency, acceptance and branching ratio and
then normalized to the number of inelastic collisions in the analyzed sample~\cite{oyama}.
Figure~\ref{fig:spectra} shows the resulting spectra, 
compared with PHOJET~\cite{PHOJET} and some widely used PYTHIA~\cite{PYTHIA} tunes: D6T~\cite{D6T},
ATLAS-CSC~\cite{ATLAS} and Perugia-0~\cite{Perugia0}.
For the $\kstar$, PYTHIA D6T shows the best agreement, while PHOJET and ATLAS-CSC produce a softer spectrum.
The $\phir$ is well reproduced by D6T below 2.5~$\gmom$ and by PHOJET and ATLAS-CSC above this limit.
Both mesons are underestimated by Perugia-0.
In contrast, all models underestimate the $\xstar$ yield.

Figure~\ref{fig:ratios} shows the $\varphi/$K$^-$ and $\textrm{K}^*/$K$^-$ ratios, for different energies and different 
colliding systems~\cite{ratios1,ratios2}.
including ALICE measurements~\cite{marek,paper900gev1,paper900gev2}.
Both ratios do not vary with $\sqrt{s}$.


\section{Conclusions}\label{sec:conclusions}

Spectral distributions $\mathrm{d}^2N/\mathrm{d}y\mathrm{d}\pT$ has been determined for $\kstar$, 
$\phir$ and $\xstar$ resonances through their hadronic decay channels.
Spectra show different levels of agreement with various event generators: mesons
are reasonably well reproduced 
(except by Perugia-0)
while the $\xstar$ is underestimated in all cases.
Ratios of $\phir$ and $\kstar$ to K$^-$ do not increase with respect to lower energies.

Other resonances studies are ongoing to complement the picture for pp collisions at 7 TeV:
$\Delta^{\pm\pm}$, $\Lambda(1520)$ and $\sstar$.
The analysis of Pb--Pb collisions at $\sqrt{s_{\mathrm{NN}}} = 2.76$~$\ten$ has also started with the extraction of $\phir$ signal 
for $0.5 \leq \pT \leq 5$~$\gmom$, both in central and peripheral events.

\end{document}